\documentclass[aps,prb,showpacs,nofootinbib,superscriptaddress,twocolumn,longbibliography]{revtex4-2}

\usepackage{amsfonts}
\usepackage{amsmath}
\usepackage{multirow}
\usepackage{txfonts}
\usepackage{amssymb}
\usepackage{amsbsy}
\usepackage{graphicx}
\usepackage{epstopdf}
\usepackage{color}
\usepackage{braket} 
\usepackage{mathdots} 
\usepackage{booktabs}
\usepackage{indentfirst}
\usepackage{hyperref}
\usepackage{diagbox}
\usepackage{lipsum}
\usepackage{mwe}
\hypersetup{ colorlinks=true, linkcolor=blue, anchorcolor=blue, citecolor=blue,urlcolor=blue}
\usepackage{cleveref}
\crefname{appendix}{App.}{Apps.}
\crefname{equation}{Eq.}{Eqs.}
\crefname{figure}{Fig.}{Figs.}
\crefname{table}{Tab.}{Tabs.}
\crefname{section}{Sec.}{Secs.}
\begin{document}
\title{From the Quantum Breakdown Model to the Lattice Gauge Theory}	

\author{Yu-Min Hu}
\affiliation{Institute for Advanced Study, Tsinghua University, Beijing,  100084, China}
\affiliation{Department of Physics, Princeton University, Princeton, NJ 08544, USA}
	
\author{Biao Lian}
\altaffiliation{biao@princeton.edu}
\affiliation{Department of Physics, Princeton University, Princeton, NJ 08544, USA}

\date{\today}\begin{abstract}
The one-dimensional quantum breakdown model, which features spatially asymmetric fermionic interactions simulating the electrical breakdown phenomenon, exhibits an exponential U(1) symmetry and a variety of dynamical phases including many-body localization and quantum chaos with quantum scar states. We investigate the minimal quantum breakdown model with the minimal number of on-site fermion orbitals required for the interaction, and identify a large number of local conserved charges in the model. We then reveal a mapping between the minimal quantum breakdown model in certain charge sectors and a quantum link model which simulates the U(1) lattice gauge theory, and show that the local conserved charges map to the gauge symmetry generators. A special charge sector of the model further maps to the PXP model, which shows quantum many-body scars. This mapping unveils the rich dynamics in different Krylov subspaces characterized by different gauge configurations in the quantum breakdown model.
\end{abstract}

\maketitle

\section{Introduction}

The study of nonequilibrium quantum dynamics in many-body systems has been a longstanding pursuit in contemporary condensed matter physics. The unitary time evolution of a generic nonintegrable quantum many-body system would approach the thermal equilibrium, a phenomenon closely associated with the eigenstate thermalization hypothesis (ETH) \cite{Deutsch1991quantum,Srednicki1994chaos,DAlessio2016from,deutsch2018eigenstate}. In recent years, extensive studies have been exploring quantum systems that violate the ETH. Notably,  the many-body localization provides an interesting possibility of ETH violation by introducing disorders \cite{Arijeet2010MBL,Znidaric2008MBL,Dmitry2019colloquium,abanin2017recent,Oganesyan2007localization,nandkishore2015many,Gornyi2005interacting}. More recently, the ETH violation due to quantum many-body scar states and Hilbert space fragmentation has also greatly attracted both theoretical and experimental investigations \cite{bernien2017probing,turner2018weak,turner2018quantum, Sala2020ergodicity, Khemani2020Localization, Yang2020Hilbert, serbyn2021quantum,chandran2023quantum,Moudgalya2022Hilbert,moudgalya2022quantum}.

Lattice gauge theory provides an alternative approach to ETH violation systems, leading to a wide class of dynamical phenomena associated with the configurations of gauge fields. In particular, the gauge degrees of freedom may induce disorder-free localization in quantum systems \cite{Smith2017disorder-free, Smith2018dynamical, Brenes2018many-body, Papaefstathiou2020disorder-free, Karpov2021disorder-free, Hart2021logarithmic, Lang2022disorder-free, Halimeh2022enhancing, Chakraborty2022disorder-free}. Besides, the lattice gauge theory can also hold quantum many-body scar states embedded in thermal eigenstates \cite{Banerjee2021quantum,Aramthottil2022scars, Yao2022quantum,su2023observation, Desaules2023prominent,Desaules2023weak, Halimeh2023robustquantummany}.

Recently, an intriguing quantum many-body system, called the quantum breakdown model, was proposed to describe the dielectric breakdown process from a microscopic perspective \cite{lian2023quantumbreakdown,liu20232d,hu2024bosonic,chen2024quantum}. The one-dimensional (1D) fermionic quantum breakdown model features a spatially asymmetric breakdown interaction that annihilates a fermion at one site and simultaneously creates more fermions at the neighboring site on the right \cite{lian2023quantumbreakdown}. With an increasing number of fermion orbitals (flavors) at each site, the quantum breakdown model undergoes a crossover from the many-body localization phase to the quantum chaotic phase with scar states. 

In this paper, we investigate the minimal quantum breakdown model, which necessitates the smallest number of fermion orbitals that are compatible with the spatially asymmetric breakdown interactions. Under this circumstance, we can identify an extensive number of local conserved quantities that contribute to the fragmentation of the total Hilbert space. Notably, we find that such a minimal quantum breakdown model in certain charge sectors can be mapped to a model of lattice gauge theory, known as the quantum link model \cite{Kogut1975hamiltonian, Brower1999QCD, chandrasekharan1997quantum, wiese2013ultracold}. The latter can be experimentally simulated in various quantum devices \cite{Banerjee2012atomic, Hauke2013quantum, Marcos2013superconducting, Rico2014tensor, Stannigel2014constrained, zohar2015quantum, Martinez2016Realtime, mil2020scalable, banuls2020simulating, Davoudi2020towards,  Surace2020lattice, yang2020observation, zhou2022thermalization, Wang2023Interrelated}. Through this mapping, the local conserved quantities in the quantum breakdown model play the role of the gauge symmetry generators in the lattice gauge theory. As a result, Krylov subspaces with distinct gauge configurations give rise to various subspace dynamics, ranging from free fermions on hypercubic lattices with boundary defects to strongly interacting sectors with quantum many-body scars. Our results reveal that the lattice gauge theory not only offers a theoretical perspective to understand the dynamics in the quantum breakdown model but also provides a practical approach to simulating this model in advanced quantum experiments.

The rest of this paper is organized as follows. In Sec.\ref{sec:Hamiltonian}, we introduce the Hamiltonian of the quantum breakdown model and identify its symmetry and conserved quantities.  Then in Sec.\ref{sec:mapping}, we map the minimal quantum breakdown model to $U(1)$ lattice gauge theory. Based on this mapping, we discuss various quantum dynamical behaviors in certain representative gauge sectors in Sec. \ref{sec:dynamics}. Our work is then concluded in Sec.\ref{sec:discussion}.

\section{Quantum breakdown model}\label{sec:Hamiltonian}
\subsection{Model Hamiltonian}
     
The breakdown process of a dielectric gas subjected to a sufficiently strong electric field can be phenomenologically described as follows. Because of the strong electric field,  the neutral atom can be ionized into one electron and one ion. Then the free electron is immediately accelerated by the strong electric field. On the contrary, the produced ion is accelerated in the opposite direction, but experiences much slower dynamics because of its much heavier mass. Therefore, we ignore the ion dynamics and focus only on the fast motions of electrons. Subsequently, the fast electrons collide with other atoms, triggering the progressive generation of additional electrons and ions. As a result, more and more electrons are generated and accelerated by the electric field, leading to a Townsend particle avalanche of electrons \cite{townsend1910theory}.

By ignoring the ions, this breakdown process can be effectively described by a microscopic Hamiltonian called the \emph{quantum breakdown model} \cite{lian2023quantumbreakdown}. We consider a 1D system with $M$ sites, each site having $N$ fermion orbitals (flavors). The generic Hamiltonian is given by
\begin{equation}
    \hat H=\hat H_I+\hat H_\mu.\label{eq:Ham}
\end{equation}
With $\hat c_{m,i}^\dagger$ and  $\hat c_{m,i}$ being the creation and annihilation operators of the $i$th fermionic mode at the $m$th site, the interacting part $\hat H_I$ represents the spatially asymmetric breakdown interaction 
\begin{equation}
    \hat H_I=\sum_{m=1}^{M-1}\sum_{l=1}^{N}\sum_{i_1<\cdots<i_{2q+1}}^{N}\left[J_{m, l}^{i_1 i_2\cdots i_{2q+1}}\left(\prod_{k=1}^{2q+1}\hat c_{m+1, i_k}^{\dagger}\right) \hat c_{m, l}+\text { h.c.}\right]\text {, }\label{eq:q-breakdown}
\end{equation}
Here, $\text{h.c.}$ is the Hermitian conjugate, and $q$ is a nonnegative integer. The values of the interaction strength $J_{m, l}^{i_1 i_2\cdots i_{2q+1}}$ are complex numbers that are arranged antisymmetrically with respect to the indices $i_1,\cdots,i_{2q+1}$. This asymmetric interaction indicates that the annihilation of one fermion leads to the creation of $2q+1$ fermions at the adjacent site. Therefore, Eq.\eqref{eq:q-breakdown} defines a class of quantum breakdown models with different $q$ values. The asymmetrical interaction is defined to maintain the fermion parity. For the nontrivial $\hat H_I$ to be valid, the number of fermion orbitals  per site must satisfy
\begin{equation}
N\ge 2q+1\ .
\end{equation}

The second part $\hat H_\mu$ is the on-site potential, which is given by
\begin{equation}\label{eq:Hmu}
    \hat H_\mu=\sum_{m=1}^M\mu_m \hat n_{m},\quad \hat n_m=\sum_{i=1}^N\hat c_{m,i}^\dagger\hat c_{m,i}.
\end{equation}
Here, $\mu_m$ represents the potential at the $m$th site. Also, $\hat n_m$ is the fermion number operator at the $m$th site.

The quantum breakdown model displays a wide variety of dynamical phases, including many-body localization, Hilbert space fragmentation, and quantum chaos \cite{lian2023quantumbreakdown,liu20232d,hu2024bosonic,chen2024quantum}. As shown in Ref.\cite{lian2023quantumbreakdown}, the $q=1$ quantum breakdown model has almost all eigenstates solvable when $N=3$, while it exhibits quantum chaos with many-body scar states when $N$ is large. Interestingly, a dynamical breakdown transition is controlled by the ratio between the interaction strength and the on-site potential. A considerably large interaction is required to overcome the energy barrier induced by the on-site potential, leading to the proliferation of electrons.  

In this paper, we study the \emph{minimal quantum breakdown models} which are defined by the requirement
\begin{equation}\label{eq-minimal-QB}
N=2q+1\ .
\end{equation}
In contrast to the quantum chaotic phase established in the large $N$ regime, we will show that the $N=2q+1$ case has a large number of conserved quantities that make this model almost exactly solvable.

\begin{figure}
    \centering
    \includegraphics[width=8cm]{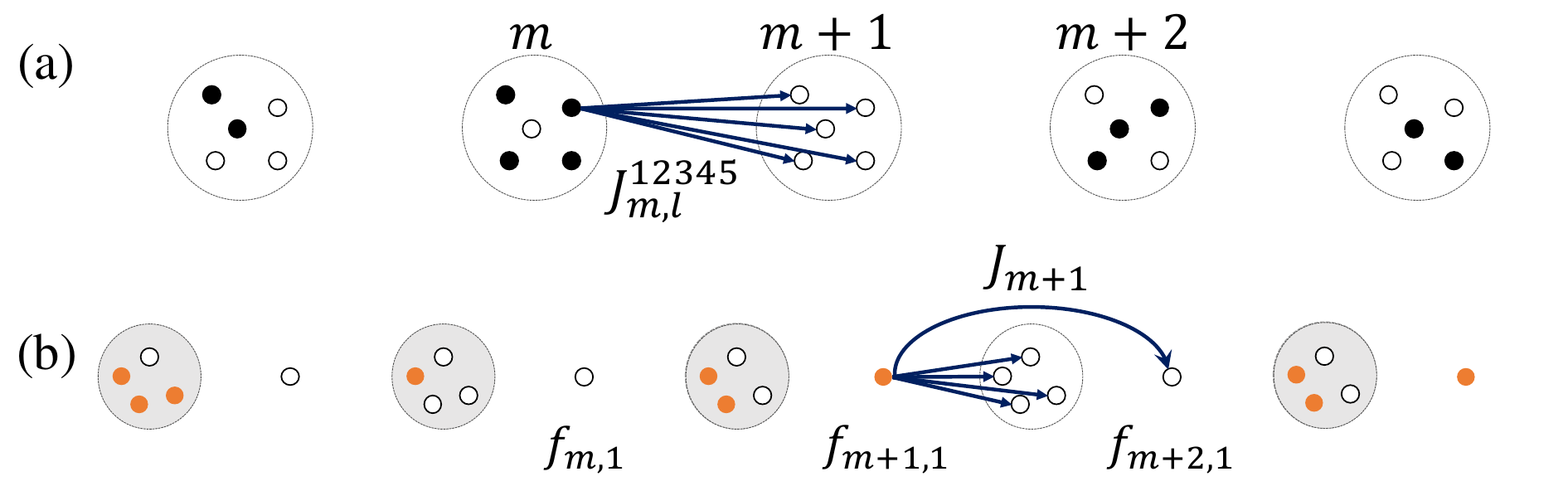}
    \caption{(a) The $q=2$ quantum breakdown model. (b) The quantum breakdown model after the basis rotation. A solid circle ($\bullet$) denotes an occupied orbital, while a hollow circle ($\circ$) indicates an unoccupied orbital.}
    \label{fig:model}
\end{figure}

\subsection{Symmetry and conserved quantity}\label{sec:symmetry}
To obtain the conserved quantities of the minimal quantum breakdown model, we need to analyze its symmetry. In this section, we focus on the global symmetries of this model. We first consider the following spatially dependent unitary transformation:
\begin{equation}
    \hat V_m\hat c_{m,i} \hat V_m^\dagger=e^{i\varphi_m}\hat c_{m,i},\quad \hat V_m=e^{-i\varphi_m\hat n_m}.
\end{equation}
Then the invariance of the breakdown Hamiltonian requires that 
\begin{equation}
    \varphi_m=(2q+1)\varphi_{m+1} \mod 2\pi.\label{eq:phase_relation}
\end{equation}
Moreover, as we will show below, the boundary conditions have a significant effect on this equation, as the periodic boundary condition (PBC) necessitates an extra restriction between the first site and the last site, which is not present under the open boundary condition (OBC).

We first discuss the symmetry of the PBC. In this case, the phase relations of $\hat V_m$ that keep the Hamiltonian invariant are given by Eq.\eqref{eq:phase_relation} for $m=1,2,\dots,(M-1)$. Assuming $\varphi_M=\varphi$ where $\varphi$ is a site-independent constant angle, these relations immediately lead to a close solution $\varphi_m=(2q+1)^{M-m}\varphi$. Furthermore, the PBC further imposes a constraint $\varphi_M=(2q+1)\varphi_1 \mod 2\pi$, which requires $\varphi$ to satisfy the following condition:  $\varphi=(2q+1)^M\varphi \mod 2\pi$. This condition implies that the angle $\varphi$ can only take discrete values
\begin{equation}
\varphi=\frac{2\pi p}{(2q+1)^M-1}\ ,\ \ \ (p=0,1,\dots,(2q+1)^M-2)
\end{equation}
Therefore, the quantum breakdown model under the PBC has a discrete $\mathbb{Z}_{(2q+1)^M-1}$ symmetry, a global symmetry depending on the system size \cite{Watanabe2023ground,delfino20232d,Hu2023spontaneous}. 

For the OBC, the phase relations in Eq.\eqref{eq:phase_relation} give rise to a solution $\varphi_m=(2q+1)^{M-m}\varphi$ with $\varphi\in[0,2\pi)$ taking continuous values. Therefore, the quantum breakdown model with OBC has a spatially modulated global $U(1)$ symmetry called the \emph{exponential symmetry} \cite{sala2022dynamics,hu2024bosonic,sala2023exotic,Hu2023spontaneous,lian2023quantumbreakdown,Watanabe2023ground,delfino20232d,han2023topological}. This exponential symmetry is generated by an exponential $U(1)$ charge
\begin{equation}
    \hat Q=\sum_{m=1}^M(2q+1)^{M-m}\hat n_m.\label{eq:c_charge}
\end{equation}
The conserved charge $\hat Q$ implies that the fermions at the $m$th site have an effective charge $q_m=(2q+1)^{M-m}$. Intuitively, the asymmetric breakdown interactions annihilate one fermion at the $m$th site and create $2q+1$ fermions at the adjoint site, splitting the effective charge into $2q+1$ pieces. This conserved charge makes it possible to use exact diagonalization to study the energy spectrum and quantum dynamics within each charge sector \cite{lian2023quantumbreakdown}. 

\subsection{Extensive conserved quantities in the minimal quantum breakdown model} 
While the symmetry analysis in Sec.\ref{sec:symmetry}  applies to the quantum breakdown model with $N\ge2q+1$, the minimal quantum breakdown model with $N=2q+1$ (\cref{eq-minimal-QB}) has a richer and more interesting structure which we will focus on in the rest of the paper.  As we shall show below, an extensive number of local conserved quantities exist in the $N=2q+1$ quantum breakdown model. These conserved quantities result in exponentially many disconnected Krylov subspaces, a hallmark of Hilbert space fragmentation. For simplicity, we impose OBC here, for which case we do not need to worry about the relation between the first site and the last site. The analysis for symmetries under PBC and the corresponding dynamical properties are however similar. 

To extract the conserved quantities, we express the minimal quantum breakdown model in a simpler form. Specifically, we employ a local $U(2q+1)$ unitary transformation among the $2q+1$ fermion flavors on the $m$th site as
\begin{equation}
    \hat c_{m,i}=\sum_{j}U_{i,j}^{(m)}\hat f_{m,j},
\end{equation}
where $U^{(m)}$ is an element in the $U(2q+1)$ group. On the one hand, the uniform on-site potential $\hat H_\mu$ in \cref{eq:Hmu} is invariant under this transformation, since $\hat n_m=\sum_{i=1}^{2q+1}\hat c_{m,i}^\dagger \hat c_{m,i}=\sum_{i=1}^{2q+1} \hat f_{m,i}^\dagger\hat f_{m,i}$. On the other hand, since there are only $2q+1$ fermion modes per site, we have $ \hat c_{m,1}\hat c_{m,2}\cdots\hat c_{m,2q+1}= \det[U^{(m)}]\hat f_{m,1}\hat f_{m,2}\cdots\hat f_{m,2q+1}$ and  $\hat c_{m,1}^\dagger \hat c_{m,2}^\dagger\cdots\hat c_{m,2q+1}^\dagger= \det[U^{(m)}]^*\hat f_{m,1}^\dagger\hat f_{m,2}^\dagger\cdots\hat f_{m,2q+1}^\dagger$, namely, $c_{m,1}\hat c_{m,2}\cdots\hat c_{m,2q+1}$ and its Hermitian conjugate transform as a singlet in the on-site flavor space. Then the breakdown interaction $\hat H_I$ transforms as
\begin{equation}
    \hat H_I=\sum_{m=1}^{M-1}\sum_{l,l'=1}^{2q+1}J_{m,l}^{12\cdots(2q+1)}[\det U^{(m+1)}]^*U_{l,l'}^{(m)}\left(\prod_{k=1}^{2q+1}\hat f_{m+1, k}^{\dagger}\right)\hat f_{m,l'}+\text{h.c.}
\end{equation}
Since the coefficients $J_{m,l}^{12\cdots(2q+1)}$ is a vector with component $l$ under the $U(2q+1)$ rotation, we can always choose the unitary transformation $U^{(m)}$ such that \cite{lian2023quantumbreakdown}
\begin{equation}
   \sum_{l=1}^{2q+1} J_{m,l}^{12\cdots(2q+1)}[\det U^{(m+1)}]^*U_{l,l'}^{(m)}=J_m\delta_{l',1}.
\end{equation}
The real number $J_m$ is the norm of the vector $(J_{m,1}^{12\cdots(2q+1)},J_{m,2}^{12\cdots(2q+1)},\cdots,J_{m,2q+1}^{12\cdots(2q+1)})^T$, which is given by
\begin{equation}
J_m=\sqrt{\sum_{l=1}^{2q+1}\left|J_{m,l}^{12\cdots(2q+1)}\right|^2}\ .
\end{equation}
Then the minimal quantum breakdown model takes a much simpler form:
\begin{equation}
\begin{aligned}
& \hat H=\hat H_I+\hat H_\mu ,\\
& \hat H_I=\sum_{m=1}^{M-1}\left[ J_m\left(\prod_{i=1}^{2q+1}\hat f_{m+1, i}^{\dagger}\right)\hat f_{m,1}+\text { h.c. }\right], \\
& \hat H_\mu=\sum_{m=1}^M \mu_m \hat{n}_m.
\end{aligned}
\label{eq:f_breakdown}
\end{equation}
After the basis transformation, only fermions on the first orbital can move between different sites. A fermion moving in the right direction generates additional $2q$ fermions at the right adjacent site [Fig.~\ref{fig:model}(b)]. Once these $2q$ fermions are created, they become immobile, incapable of further moving rightward. The only possible dynamics for these $2q$ fermions is their simultaneous annihilation when a fermion on the first orbital moves to the left site, namely, the Hermitian conjugate of their creation process. We note that such a simplification of Hamiltonian similar to \cref{eq:f_breakdown} via local unitary transformations is not applicable to generic quantum breakdown models with $N>2q+1$. 

The simplified form of the minimal quantum breakdown model in \cref{eq:f_breakdown} allows us to reveal many more hidden conserved charges. To see this, we attach an effective charge, denoted as $q_{m,i}$, to the $i$th orbital $f$-fermion on the $m$th site. Consequently, it is straightforward to see that the following modulated charge $\hat Q(\{q_{m,i}\})$ commutes with the Hamiltonian $\hat H$ in Eq.~\eqref{eq:f_breakdown} provided that $q_{m,i}$ satisfy the following condition:
\begin{equation}
    \hat Q(\{q_{m,i}\})=\sum_{m=1}^M\sum_{i=1}^{2q+1}q_{m,i}f_{m,i}^\dagger f_{m,i}\ ,\quad q_{m,1} = \sum_{i=1}^{2q+1} q_{m+1,i}\ .\label{eq:f_general_charge}
\end{equation}
The above charge $\hat Q(\{q_{m,i}\})$ in Eq. \eqref{eq:f_general_charge} reduces to the conserved charge $\hat Q$ for $c$-fermions in Eq.~\eqref{eq:c_charge} if one chooses $q_{m,i} = (2q+1)^{M-m}$. Clearly, the arbitrariness of $q_{m,i}$ in \cref{eq:f_general_charge} gives rise to many more conserved quantities.
 
To further extract the conserved quantities encoded in Eq.~\eqref{eq:f_general_charge}, we can reformulate the charge constraint as $\sum_{i=1}^{2q+1} q_{m+1,i}/q_{m,1} = 1$. In particular, we choose the ratios $q_{m+1,i}/q_{m,1}$ to be given by the following parameters:
\begin{equation}
    \begin{split}
        q_{m+1,1}/q_{m,1}&=\gamma,\\
        q_{m+1,2}/q_{m,1}&=(1-\gamma)/(2q)+\beta_2,\\
        q_{m+1,3}/q_{m,1}&=(1-\gamma)/(2q)+\beta_3-\beta_2,\\
        \cdots\\
        q_{m+1,i}/q_{m,1}&=(1-\gamma)/(2q)+\beta_{i}-\beta_{i-1},\\
         \cdots\\
        q_{m+1,2q}/q_{m,1}&=(1-\gamma)/(2q)+\beta_{2q}-\beta_{2q-1},\\
        q_{m+1,2q+1}/q_{m,1}&=(1-\gamma)/(2q)-\beta_{2q}.
    \end{split}
\end{equation}

Here, $\gamma$ and $\beta_i$ with $i=2,3,\dots,2q$ represent the specific $2q$ free parameters (which can be any complex numbers). Consequently, the conserved quantity in Eq.~\eqref{eq:f_general_charge} transforms into:

\begin{equation}
\begin{split}
    \hat Q(\{q_{m,i}\})=&\sum_{i=2}^{2q+1}q_{1,i}\hat f_{1,i}^\dagger\hat  f_{1,i}+q_{1,1}\sum_{m=1}^M\gamma^{m-1} \hat f_{m,1}^\dagger\hat  f_{m,1}\\
    &+q_{1,1}\sum_{m=2}^M\gamma^{m-2}\frac{1-\gamma}{2q}\sum_{i=2}^{2q+1}\hat f_{m,i}^\dagger \hat f_{m,i}\\
   & +q_{1,1}\sum_{m=2}^M\gamma^{m-2} \sum_{i=2}^{2q}\beta_i(\hat f_{m,i}^\dagger \hat f_{m,i}-\hat f_{m,i+1}^\dagger\hat  f_{m,i+1}).
    \end{split}
\end{equation}
$q_{1,i}$ for $i=1,\cdots,2q+1$ represent the effective charges for fermions at the first site. The conservation of the total charge is independent of the choices of the free parameters $\gamma$ and $\beta_i$, thus the coefficient of each power of these free parameters can be identified as an independent conserved quantity. All of these coefficients give a large number of local conserved quantities. Notably, these local conserved quantities can be categorized into three distinct sets as follows.

The first set of conserved quantities is localized on the first site and is denoted as:
\begin{equation}\label{eq:Qa_Breakdown}
    \hat Q_{a,i} = \hat f_{1,i}^\dagger \hat f_{1,i}, \quad i = 2, \cdots, 2q+1,
\end{equation}
This set signifies that the fermions on $2q$ orbitals of the first site are entirely decoupled from the rest of the system. These $2q$ conserved quantities at the first site directly stem from taking the OBC.

The second set comprises on-site conserved quantities as:
\begin{equation}\label{eq:Qb_Breakdown}
    \hat Q_{b,m,i} = \hat f_{m,i}^\dagger \hat f_{m,i} - \hat f_{m,i+1}^\dagger \hat f_{m,i+1},
\end{equation}
where the site index $m = 2,\dots,M$ and orbital index $i = 2,\dots,2q$. These on-site conserved quantities indicate that the $\hat f_{m,i>1}$ fermions at the $m$th site are subject to simultaneous annihilation and creation, revealing the conservation of the population imbalance between $f$-fermion orbitals with orbital indices $i>1$. This behavior aligns directly with the breakdown interaction in Eq.~\eqref{eq:f_breakdown}.

The third set of conserved quantities is also local, but intriguingly not on-site, which is given by:
\begin{equation}\label{eq:Qc_Breakdown}
    \hat Q_{c,m} = \hat f_{m,1}^\dagger \hat f_{m,1} + \frac{1}{2q}\sum_{i=2}^{2q+1}\left(\eta_m\hat f_{m+1,i}^\dagger \hat f_{m+1,i} - \tilde\eta_m\hat f_{m,i}^\dagger \hat f_{m,i}\right),
\end{equation}
where $m = 1,\dots,M$. These conserved quantities describe the interactions between the first fermion $\hat f_{m,1}$ at each site and the other fermions. We impose the coefficients $\eta_m = 1-\delta_{m,M}$ and $\tilde\eta_m = 1-\delta_{m,1}$ such that these conserved quantities are compatible with the OBC.

As a result, we obtain an extensive number of local conserved operators for the minimal quantum breakdown Hamiltonian in Eq.\eqref{eq:f_breakdown}. These operators in \cref{eq:Qa_Breakdown,eq:Qb_Breakdown,eq:Qc_Breakdown} can further add or multiply to generate additional conserved quantities, serving as generators of the underlying commutant algebra \cite{moudgalya2022quantum}. For example, the total fermion number $\hat{\mathcal{N}}_1 = \sum_{m=1}^M \hat f_{m,1}^\dagger \hat f_{m,1}$ on the first orbital of all sites is conserved, which is equivalent to $\hat{\mathcal{N}}_1 = \sum_{m=1}^M \hat Q_{c,m}$.

As implied by the conserved quantities $\hat Q_{a,i}$, the fermions created by $\hat f_{1,i}^\dagger $ with $i>1$ remain frozen in their dynamics. However, the remaining two groups of conserved quantities, $\hat Q_{b,m,i}$ and $\hat Q_{c,m}$, play a crucial role in shaping the connected Hilbert subspaces (i.e., Krylov subspaces) and, thereby, constraining the quantum dynamics of the minimal quantum breakdown model with $N=2q+1$.

The breakdown interactions in Eq.~\eqref{eq:f_breakdown} simultaneously annihilate or create all fermions with orbital index $i>1$. Consequently, the subspace that exhibits nontrivial dynamics at the $m$th site must exclusively comprise states with an identical number of fermions with orbital index $i>1$. Conversely, a site with an uneven distribution of these fermion modes becomes dynamically frozen. These two distinct types of states can be distinguished by the eigenvalues of the second set of conserved charges $\hat Q_{b,m,i}$. In a subspace where any $\hat Q_{b, m, i}$ has a nonzero eigenvalue, the fermions at the $m$th site with orbital index $i>1$ will remain static. As a result, this site serves as a \emph{blocking site} \cite{lian2023quantumbreakdown}, effectively dividing the system into two dynamically isolated regions. This dynamical constraint precisely exemplifies the Hilbert space fragmentation in the $N=2q+1$ quantum breakdown model.

Assume that there are two blocking sites at $m_1$ and $m_2$ and no other blocking site in between. The nontrivial dynamics then exist in the region $m_1<m< m_2$, in which any $\hat Q_{b, m, i}$ with $m_1+1\le m\le m_2-1$ and $2\le i\le 2q+1$ would have a zero eigenvalue for the states in this subspace. Within such a subspace, the quantum dynamics is further influenced by the conserved charges $\hat Q_{c,m}$. In the subsequent section, we will demonstrate that the minimal quantum breakdown model in certain subspaces is equivalent to a $U(1)$ lattice gauge model. In the language of lattice gauge theory, the conserved quantities $\hat Q_{c,m}$ play the role of gauge symmetry generators.

\section{Mapping to the lattice gauge theory}\label{sec:mapping}

Without loss of generality, we make a slight change of notation in the following discussion. Here, we assume $m_1=1$ and $m_2=M+1$ are two blocking sites, between which the region is connected without any other blocking sites. As explained in the above section, the states $\ket{\psi}$ in this dynamical subspace are restricted by
\begin{equation}
    \hat Q_{b,m,i}\ket{\psi}=0, \quad 2\le m\le M,\ i=2,\cdots,2q+1.\label{eq:zero_condition}
\end{equation}

To further simplify the minimal quantum breakdown model in Eq. \eqref{eq:f_breakdown}, we define a set of operators as
\begin{equation}
\begin{split}
    \hat F_m^\dagger&=\hat f_{m,2}^\dagger\hat f_{m,3}^\dagger\cdots \hat f_{m,2q}^\dagger\hat f_{m,2q+1}^\dagger,\\
    \hat F_m&=\hat f_{m,2q+1}\hat f_{m,2q}\cdots \hat f_{m,3}\hat f_{m,2}, \\
    \hat N_m&=\frac{1}{2q}\sum_{i=2}^{2q+1}\hat f_{m,i}^\dagger \hat f_{m,i}.
\end{split}
\end{equation}
$\hat F_m$ and $\hat F_m^\dagger$ are the operators for collectively annihilating and creating $2q$ fermions at the $m$th site. Then the quantum breakdown model in Eq.\eqref{eq:f_breakdown} can be re-expressed as
\begin{equation}
 \hat   H=\sum_{m=1}^{M-1}( J_m\hat f_{m+1,1}^\dagger \hat F_{m+1}^\dagger\hat f_{m,1}+\text { h.c.})+\sum_{m=1}^M \mu_m(\hat f_{m,1}^\dagger\hat f_{m,1}+2q\hat N_m).
\end{equation}
Since the breakdown interaction preserves the fermion parity, the two operators, $\hat F_m$ and $\hat F_m^\dagger$, behave like bosonic operators. They satisfy the following commutation relations:
\begin{equation}
\begin{split}
   & [\hat F_m,\hat f_{n,1}]=[\hat F_m^\dagger,\hat f_{n,1}^\dagger]=0, \\
    &[\hat F_m,\hat F_n]=[\hat F_m^\dagger,\hat F_n^\dagger]=0, \\
    &[\hat F_n^\dagger,\hat F_m]=\delta_{mn}\left(\prod_{i=2}^{2q+1}\hat f_{m,i}^\dagger\hat f_{m,i}-\prod_{i=2}^{2q+1}(1-\hat f_{m,i}^\dagger\hat f_{m,i})\right). 
\end{split}
\end{equation}

\begin{figure}
    \centering
    \includegraphics[width=8cm]{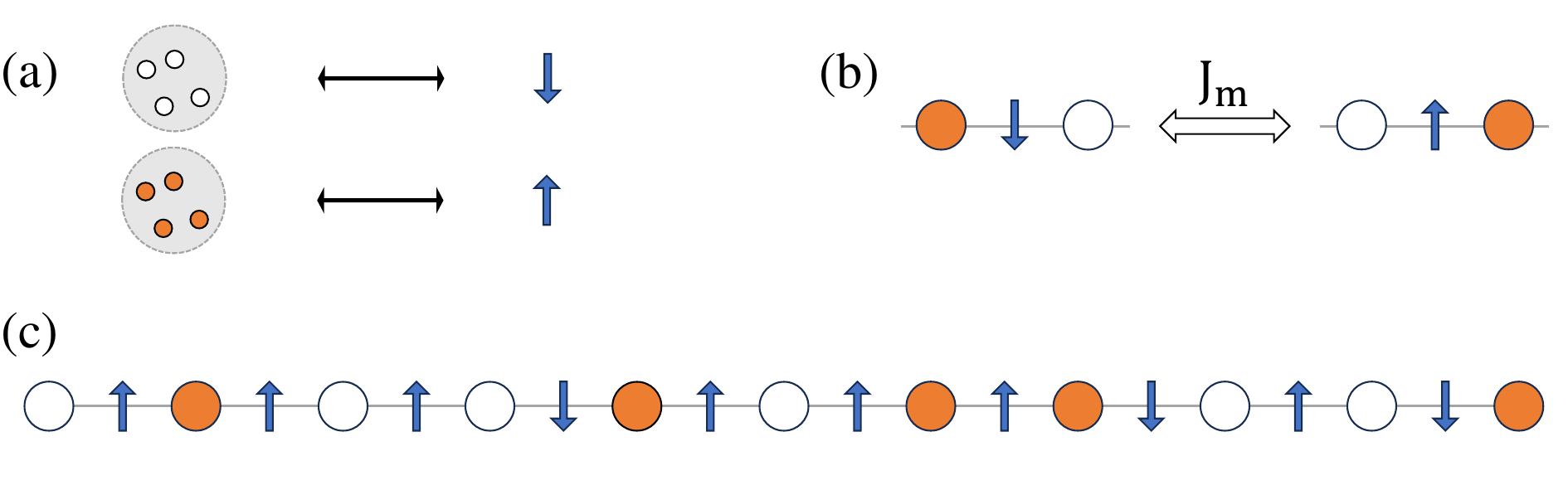}
    \caption{(a) The correspondence between fermions and link spins established in Eq. \eqref{eq:F_spin_corresp}. (b) The sketch of interactions in the minimal quantum breakdown model in Eq. \eqref{eq:link_Ham}. (c) A typical configuration in the minimal quantum breakdown model in Eq. \eqref{eq:link_Ham}. }
    \label{fig:duality}
\end{figure}

Upon imposing the constraint in Eq.~\eqref{eq:zero_condition} on the Hilbert space, only two configurations for the $2q$ immobile $f$-fermion modes (with orbital indices $2\le i\le 2q+1$) at each site are accessible: the vacuum state $\ket{\downarrow}_m\equiv\ket{0}$ and the fully occupied state $\ket{\uparrow}_m\equiv\hat F_m^\dagger\ket{0}$ [Fig.~\ref{fig:duality}(a)]. Consequently, it is straightforward to show that $[\hat F_m^\dagger,\hat F_m]\ket{\downarrow}_m=-\ket{\downarrow}_m$ and $[\hat F_m^\dagger,\hat F_m]\ket{\uparrow}_m=\ket{\uparrow}_m$. Therefore, the commutator $[\hat F_m^\dagger,\hat F_m]$ can be expressed as $[\hat F_m^\dagger,\hat F_m]_c= 2\hat N_m-1$, where the subscript $c$ denotes the subspace spanned by $\ket{\uparrow}_m$ and $\ket{\downarrow}_m$. Furthermore, $[\hat F_m,2\hat N_m-1]_c=2\hat F_m$ and $[\hat F_m^\dagger,2\hat N_m-1]_c=-2\hat F_m^\dagger$. As a result, the operators $\hat F_m$, $\hat F_m^\dagger$, and $2\hat N_m-1$ constrained in this subspace follow the same algebra as the Pauli matrices $\hat\sigma_{m,m+1}^{\pm,z}$, with the correspondence 
\begin{equation}\label{eq:F_spin_corresp}
\hat F_{m+1}\to \hat\sigma_{m,m+1}^-\ ,\ \hat F_{m+1}^\dagger\to \hat\sigma_{m,m+1}^+\ ,\ 2\hat N_{m+1}-1\to \hat\sigma_{m,m+1}^z\ .
\end{equation}
We emphasize that such an identification between fermionic operators and spin operators is exclusively applicable within the constrained Hilbert space where Eq.~\eqref{eq:zero_condition} is satisfied.

After mapping to spin operators, we can express the minimal quantum breakdown model in the subspace of Eq.~\eqref{eq:zero_condition} into the following form:
\begin{equation}
\begin{split}
   \hat H=&\sum_{m=1}^{M-1}( J_m\hat f_{m+1,1}^\dagger \hat\sigma_{m,m+1}^+ \hat f_{m,1}+\text {h.c.})\\
    &+\sum_{m=1}^M \mu_m\hat f_{m,1}^\dagger\hat f_{m,1}+\sum_{m=1}^{M-1}q\mu_m \hat\sigma_{m,m+1}^z.\label{eq:link_Ham}
\end{split}
\end{equation}
Here, we ignore the decoupled fermions on the first site with orbital indices $2\le i\le 2q+1$ and omit some constant terms. The Hamiltonian, as shown in Fig.~\ref{fig:duality}, resembles the quantum link model, which characterizes the motion of fermions along the lattice sites while simultaneously interacting with the spins on the lattice links \cite{Kogut1975hamiltonian, Brower1999QCD, chandrasekharan1997quantum, wiese2013ultracold}.  This is a lattice version of 1+1D quantum electrodynamics. Correspondingly, the local conserved quantities $\hat Q_{c,m}$ in \cref{eq:Qc_Breakdown} are transformed into 
\begin{equation}\label{eq:gauge_link}
    \hat G_m=\hat f_{m,1}^\dagger \hat f_{m,1}+\frac{1}{2}[\eta_m(\hat\sigma_{m,m+1}^z+1)-\tilde\eta_m(\hat\sigma_{m-1,m}^z+1)].
\end{equation}

The quantum link model bridges the minimal quantum breakdown model with the 1D lattice gauge theory. To see this, we introduce the Hamiltonian of the $U(1)$ lattice gauge theory\cite{wiese2013ultracold,Banerjee2012atomic}.  Here, we focus on the open boundary conditions, and the quantum link formulation of the $U(1)$ lattice gauge theory is
\begin{equation}
\begin{aligned}
   \hat{ H}_{\text{LGT}}=&-t\sum_{m=1}^{M-1}(\hat \Psi_{m}^\dagger \hat U_{m,m+1}  \hat \Psi_{m+1}+\text{h.c.})\\
    &+\mu\sum_{m=1}^M(-1)^m \hat \Psi_m^\dagger \hat \Psi_m+g\sum_{m=1}^{M-1}\hat E_{m,m+1}^2.
\end{aligned}\label{eq:U1_lattice}
\end{equation}
In this context, $\hat \Psi_m$ and $\hat \Psi_m^\dagger$ denote the fermionic annihilation and creation operators on the lattice sites. Meanwhile, $\hat E_{m,m+1}$ represents the electric field operator on the lattice links, and $\hat U_{m,m+1}=e^{i\hat A_{m,m+1}}$ corresponds to a parallel transport operator induced by the $U(1)$ link gauge field $\hat A_{m,m+1}$. They adhere to the relation $[\hat E_{m,m+1}, \hat U_{m,m+1}]=\hat U_{m,m+1}$. This relationship arises from the fact that the electric field operator $\hat E_{m,m+1}$ on the link is the canonical momentum of the link gauge field $\hat A_{m,m+1}$.

The first term in Eq.~\eqref{eq:U1_lattice} signifies the couplings between the fermions (matter fields) at the lattice sites and the gauge fields at the lattice links. The second term represents the fermion mass, while the final term characterizes the energy of electric fields. Notably, the $U(1)$ gauge symmetry of this lattice model under open boundary conditions is generated by \cite{wiese2013ultracold,Banerjee2012atomic}:
\begin{equation}\label{eq:gauge_lattice}
    \hat {\mathcal{G}}_m=\hat \Psi_m^\dagger\hat\Psi_m+\tilde\eta_m\hat E_{m-1,m}- \eta_m\hat E_{m,m+1}.
\end{equation}
These gauge generators satisfy $[\hat H_{\text{LGT}},\hat{\mathcal{G}}_m]=0$. In the lattice gauge theory for staggered fermions, the physical states in the gauge-invariant subspace satisfy $(\hat {\mathcal{G}}_m-\frac{1-(-1)^m}{2})\ket{\psi}=0$, which is a lattice manifestation of Gauss law $\triangledown\cdot E=\rho$ \cite{wiese2013ultracold, Banerjee2012atomic, Hauke2013quantum}.  

The infinite local Hilbert space dimension associated with the gauge fields on the links poses a challenge in simulating the $U(1)$ lattice gauge theory in experiments. A truncated local Hilbert space with small $\hat E_{m,m+1}^2$ eigenvalues is often adopted, which is more experimentally accessible and is justified for large coupling strength $g$ in $\hat H_{\text{LGT}}$. This truncation leads to the quantum link model, achieved by the substitutions: $\hat U_{m,m+1}\to \hat S_{m,m+1}^+$, $\hat U_{m,m+1}^\dagger\to \hat S_{m,m+1}^-$, and $\hat E_{m,m+1}\to \hat S_{m,m+1}^z$, where $S_{m,m+1}^\mu$ are spin-$S$ operators on the link. Now, the gauge fields on the links are effectively substituted by a spin $S$. Consequently, the spin-$S$ quantum link model can be expressed as \cite{wiese2013ultracold,Banerjee2012atomic} :
\begin{equation}
  \begin{aligned}
    \hat{H}_{S}=&-t^\prime\sum_{m=1}^{M-1}(\hat \Psi_{m}^\dagger \hat S_{m,m+1} ^+ \hat \Psi_{m+1}+\text{h.c.})\\
    &+\mu\sum_{m=1}^M(-1)^m \hat \Psi_m^\dagger \hat \Psi_m+g^\prime\sum_{m=1}^{M-1}(\hat S_{m,m+1}^z)^2.
\end{aligned}\label{eq:spin_QLM}
\end{equation}
The precise form of the renormalized constants $t^\prime$ and $g^\prime$ is irrelevant in our  discussion. In the case of the spin $S=\frac{1}{2}$ quantum link model, the energy term of electric fields is a constant energy offset that can hence be  dropped. When we map the $\Psi$-fermions to the $f$-fermions, the model in Eq.~\eqref{eq:spin_QLM} shares the same interaction terms as in Eq.~\eqref{eq:link_Ham} by the following substitutions: $\hat S_{m,m+1}^{\pm} \to \hat \sigma_{m,m+1}^{\mp}$ and $2\hat S_{m,m+1}^z \to -\hat \sigma_{m,m+1}^z$. It is worth noting that we have chosen a different basis for the link spins in Eq.~\eqref{eq:link_Ham} to establish a connection between the spin-up state in the quantum link model and the occupied state $\hat F_m^\dagger\ket{0}$ in the minimal quantum breakdown model.

By employing the above substitutions and setting the parameters $J_m=-t^\prime$ and $\mu_m=(-1)^m\mu$, the minimal quantum breakdown model in Eq. \eqref{eq:link_Ham} can be mapped into the $S=\frac{1}{2}$ version of the quantum link model in Eq. \eqref{eq:spin_QLM}. However, this mapping is not exact. The distinction arises from the staggered chemical potential in the minimal quantum breakdown model, which results in a staggered magnetic field for link spins in Eq.~\eqref{eq:link_Ham}.  The staggered magnetic field for link spins is different from the energy term of electric fields $g^\prime(\hat S_{m,m+1}^z)^2$ presented in Eq.\eqref{eq:spin_QLM}. An interesting future direction is to investigate the effect of this effective magnetic field on the quantum dynamics of the underlying many-body system. Nevertheless, we emphasize that the minimal quantum breakdown model shares the same gauge symmetry structure as the $U(1)$ lattice gauge theory, which can be seen in Eq.\eqref{eq:Qc_Breakdown}, Eq.\eqref{eq:gauge_link}, and Eq.\eqref{eq:gauge_lattice}.

In general, quantum simulations of the lattice gauge theory make great efforts to enforce the gauge condition $(\hat {\mathcal{G}}_m-\frac{1-(-1)^m}{2})\ket{\psi}=0$ in experiments. However, the connection between lattice gauge theory and the quantum breakdown model we revealed here suggests that the other gauge sectors may also unveil intriguing physical phenomena. In the next section, we focus on the Hamiltonian Eq.~\eqref{eq:link_Ham} and delve into several typical gauge sectors, to illustrate the rich dynamical phenomena exhibited by the minimal quantum breakdown model.

\section{Dynamically connected subspaces}\label{sec:dynamics}
In the $N=2q+1$ minimal quantum breakdown model, the existence of local conserved quantities effectively partitions the entire Hilbert space into an exponential number of disconnected Krylov subspaces. Even after resolving the conserved quantities $\hat Q_{a,i}$ and $\hat Q_{b,m,i}$, it is still possible to make further fragmentations within the Hilbert subspace constrained by $\hat Q_{b,m,i}\ket{\psi}=0$. As a result, the conserved quantities $\hat Q_{c,m}$ play an indispensable role in determining the dynamically connected subspaces. To make it clear, we employ the minimal quantum breakdown model in the form of \cref{eq:link_Ham} to illustrate the dynamical structures in the original quantum breakdown model.

\subsection{Blocking gauge configuration}

\begin{figure}
    \centering
    \includegraphics[width=8cm]{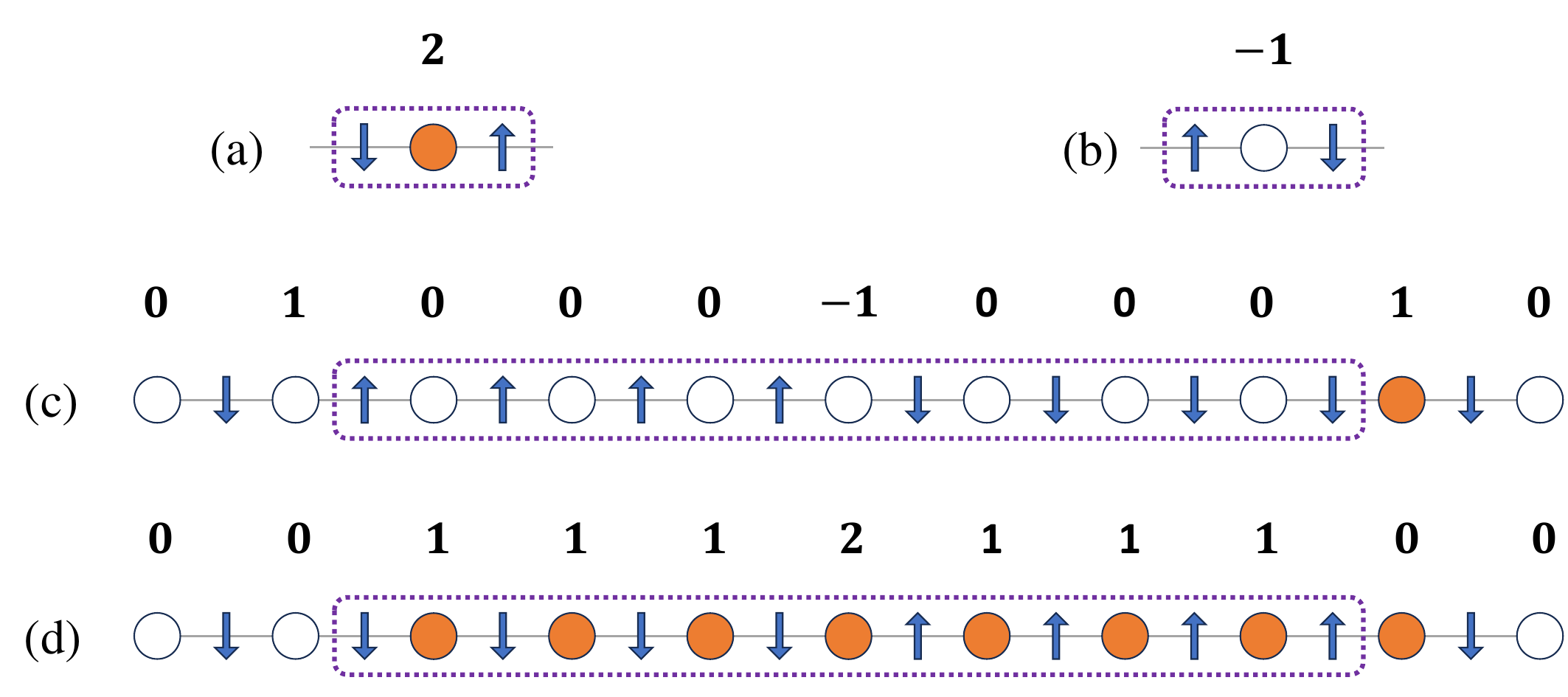}
    \caption{Different blocking configurations in the minimal quantum breakdown model in Eq. \eqref{eq:link_Ham}.}
    \label{fig:blocking}
\end{figure}

In addition to the aforementioned blocking sites with at least one $\hat Q_{b,m,i}$ satisfying $\hat Q_{b,m,i}\ket{\psi}\ne0$, there also exist two types of blocking configurations in the subspaces with all $\hat Q_{b,m,i}\ket{\psi}=0$. For simplicity, we take the form of the minimal quantum breakdown model shown in \cref{eq:link_Ham}. Such blocking configurations are determined by the local gauge generators $\hat G_m$ (i.e., $\hat Q_{c,m}$) in \cref{eq:gauge_link}, which can have eigenvalues $-1,0,1,2$ (if $m\neq 1$ or $M$). As shown in Fig.\ref{fig:blocking}, when the eigenvalue of $\hat G_m$ at the $m$th site equals $2$, the $m$th fermionic site and two surrounding spin sites admit the following configuration $\downarrow \bullet\uparrow$; similarly, when the eigenvalue of $\hat G_m$ at the $m$th site equals $-1$, we obtain a configuration $\uparrow\circ\downarrow$. \footnote{Here we use $\bullet$ and $\circ$ to represent the occupied and unoccupied fermionic sites;  $\uparrow$ and $\downarrow$ show the direction of link spins. } These two configurations are dynamically frozen. Other fermions cannot jump into this site and change the spin configuration on the nearby links. Therefore, the gauge configurations with the eigenvalue of bulk $\hat G_m$ being $2$ or $-1$ further subdivide the lattice into spatially disconnected parts.  

Furthermore, these two blocking configurations can generate more complicated blocking structures. As shown in Figs.\ref{fig:blocking}(c) and (d), the blocking configurations $\downarrow \bullet\uparrow$ and $\uparrow \circ\downarrow$ can be further extended by adding more parallel spins on the two sides. These extended blocking configurations form a domain-wall structure of the link spins, while the fermionic sites within the blocking region are either fully occupied or empty.

The existence of many blocking configurations further reduces the dimension of connected Hilbert subspaces. In the following, we only need to focus on the situations where the eigenvalues of all the bulk $\hat G_m$ are $0$ or $1$. In other words, we are free to remove the blocking region because of their frozen dynamics. Then the dynamically connected region consists of the following configurations \footnote{A slight modification is necessary for the available eigenvalues of $\hat G_m$ at the boundary sites because of the absence of link spins outside of the system. For $m=1$, the available configurations are $\circ\uparrow$ and $\bullet\downarrow$ with $\braket{\hat G_1}=$ 1. For $m=M$, the available configurations are $\downarrow\circ$ and $\uparrow\bullet$ with $\braket{\hat G_M}=$ 0.}:
\begin{equation}\label{eq:config}
    \begin{aligned}
        &\braket{\hat G_m}=0: \quad \downarrow\circ\downarrow,\ \uparrow\circ\uparrow,\ \uparrow\bullet\downarrow;\\
        &\braket{\hat G_m}=1: \quad \downarrow\bullet\downarrow,\ \uparrow\bullet\uparrow,\ \downarrow\circ\uparrow.
    \end{aligned}
\end{equation}
Here, $\braket{\hat G_m}$ represents the expectation value of $\hat G_m$ over a product state, which is also an eigenstate due to the diagonal structure of $\hat G_m$. Moreover, the combined conserved quantity $\hat G=\sum_m \hat G_m$ corresponds to the total number of fermions in the quantum link model. This quantity equals to the number of fermions $\hat {\mathcal{N}}_1$ on the first orbitals in the quantum breakdown model. Consequently, we can investigate the dynamically connected Krylov subspaces labeled with different fermion numbers, or fermion filling factors, on the first orbitals. In the following discussion about the subspace dynamics, we restrict ourselves into the Krylov subspaces below the half-filling, since the Krylov subspaces above half-filling can be readily generated via a particle-hole transformation.

\subsection{Free-fermion sector}
\begin{figure}
    \centering
    \includegraphics[width=8cm]{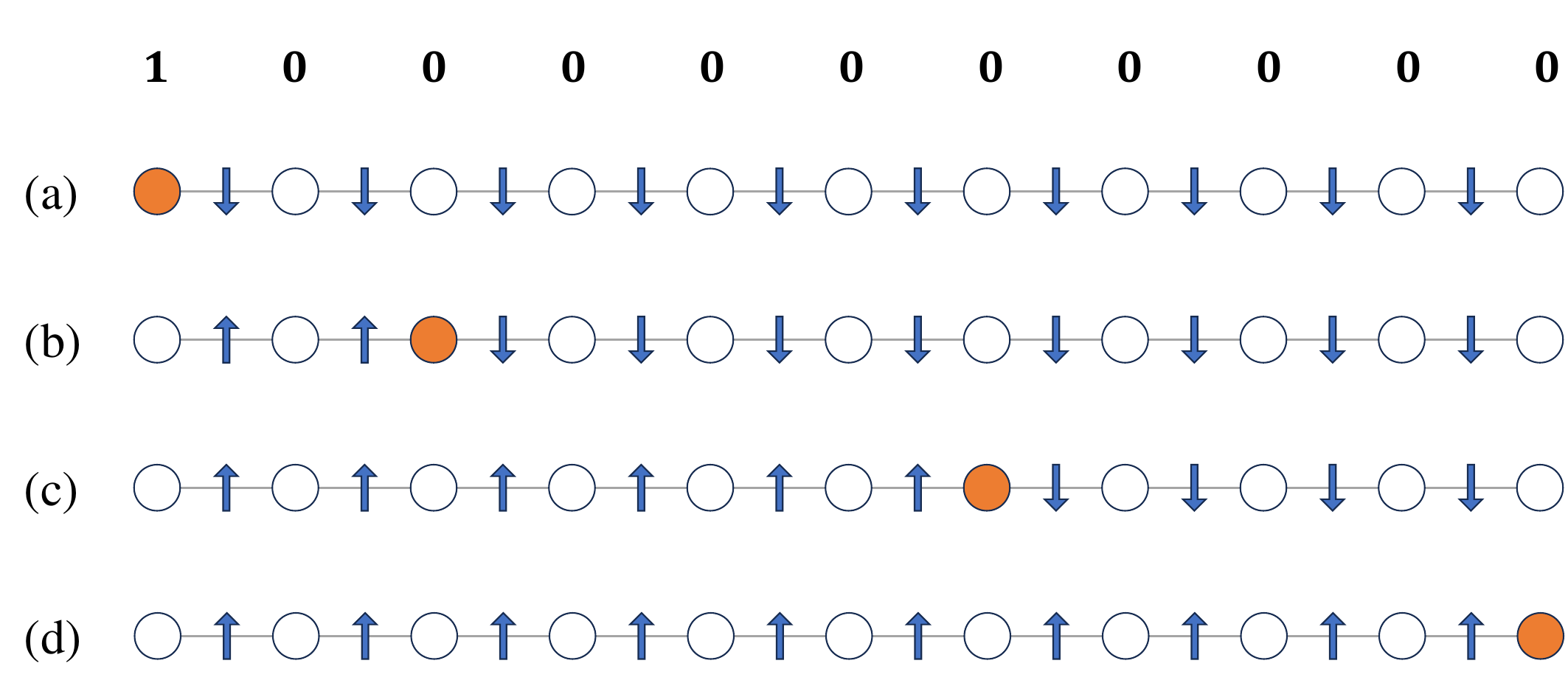}
    \caption{Typical configurations in a Krylov subspace with one fermion. The integers above show the eigenvalues of local conserved $\hat G_m$.}
    \label{fig:free}
\end{figure}
 The simplest situation is the case with only one fermion, namely, the eigenvalue of $\hat G=\sum_m \hat G_m$ is equal to $1$. Under this circumstance, the connected Krylov subspace is generated by a reference state like Fig.\ref{fig:free}(a). If we label the state by the position of the fermion as $\ket{m}$, such a state in the original quantum breakdown model can be expressed as
\begin{equation}
    \ket{m}=\hat f_{m,1}^\dagger\prod_{j=2}^{m}\left(\prod_{i=2}^{2q+1}\hat f_{j,i}^\dagger\right)\ket{0}=\hat f_{m,1}^\dagger\prod_{j=2}^{m}\hat{F}_j^\dagger\ket{0}.
\end{equation}
The reduced Hamiltonian in this Krylov subspace admits the following form:
\begin{equation}
    H_{\text{free}}=\sum_{m=1}^{M-1}(J_m\ket{m+1}\bra{m}+\text{h.c.})+\sum_{m=1}^MV_m\ket{m}\bra{m},
\end{equation}
where the onsite potential is given by $V_m=\mu_m+\sum_{i=2}^{m}2q\mu_i$. This is thus effectively a single-particle tight-binding model \cite{lian2023quantumbreakdown}.

Notably, by tuning the parameters $J_m$ and $\mu_m$, this model can exhibit different behaviors. For example, if we take a uniform $J_m=J$ and set $\mu_m=0$, this subspace describes a tight-binding model whose eigenstates are Bloch waves:
\begin{equation}
    \ket{\psi_k}=\sqrt{\frac{2}{M+1}}\sum_{m=1}^{M}\sin\left(\frac{km\pi}{M+1}\right)\ket{m},\quad k=1,\dots,M.
\end{equation}
Consequently, the eigenvalues are given by $E_k=2J\cos(\frac{k\pi}{M+1})$.  

Another interesting setup is to take a constant on-site potential $\mu_m=\mu$. In this case, $V_m=(2q(m-1)+1)\mu$ becomes a linear potential, resulting in the Wannier-Stark localization of the free fermion. Furthermore, if $J_m$ and $V_m$ are taken from some random distributions, the system will display the Anderson localization. The localization phenomenon in this subspace indicates that the original quantum breakdown model displays many-body localization in certain charge sectors.

\begin{figure}[t]
    \centering
    \includegraphics[width=8cm]{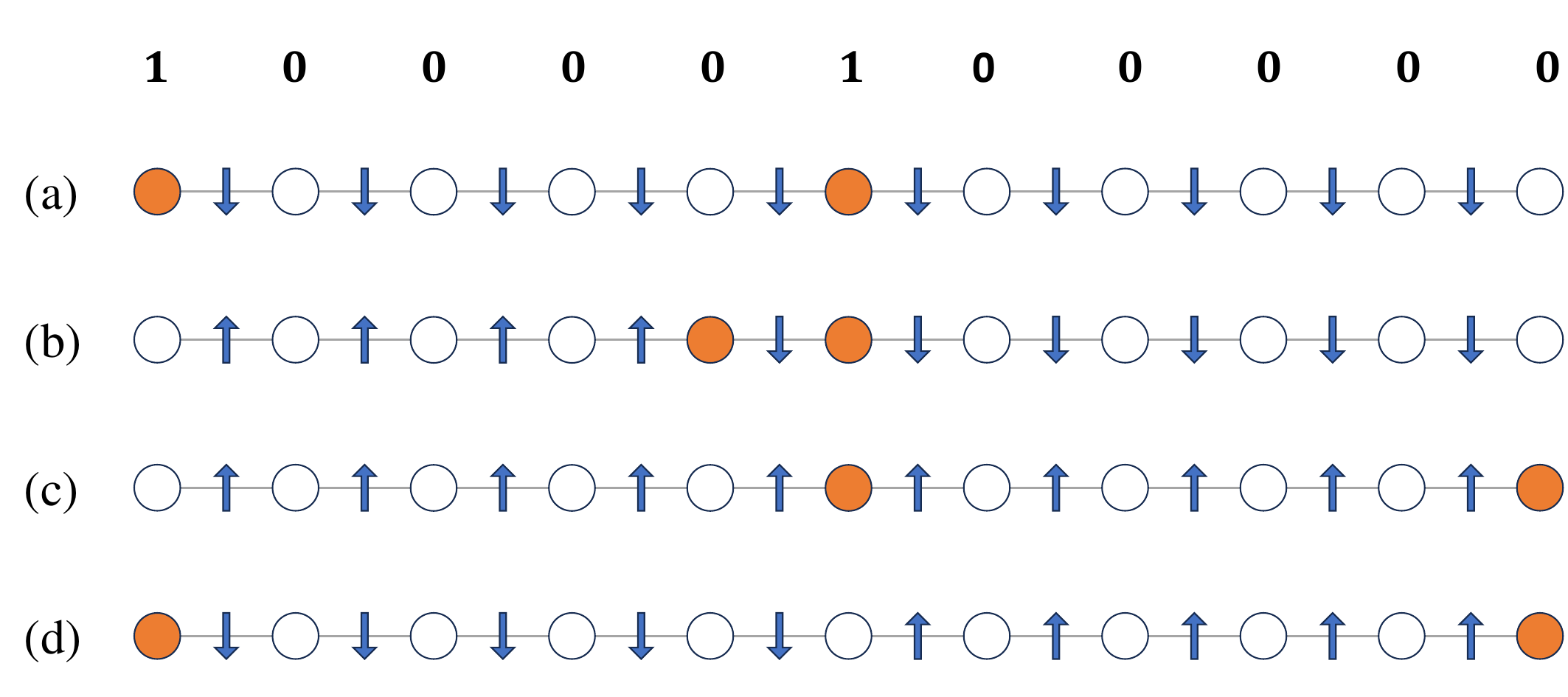}
    \caption{Typical configurations in a Krylov subspace with two fermions. The integers above show the eigenvalues of local conserved $\hat G_m$.}
    \label{fig:twoparticle}
\end{figure}

\subsection{Boundary interaction}
We now analyze Krylov subspaces that involve more than one fermion. In the subspaces with two fermions, two sites are specified with their eigenvalue of $\hat{G}_m$ being 1. A representative configuration is shown in Fig. \ref{fig:twoparticle}. In this scenario, both particles move freely within their individual dynamical zones, with the exception of a contact interaction occurring near the boundary that separates them. 

For example,  the dynamical region of the first fermion in Fig. \ref{fig:twoparticle}(a) contains the leftmost six sites, while that of the second fermion comprises the rightmost six sites. Starting from the reference state in Fig. \ref{fig:twoparticle}(a), if the second particle moves to the right, the first particle can migrate to the initial site of the second particle [Fig. \ref{fig:twoparticle}(c)]. However, due to the constraint imposed by the spin configurations, the first particle cannot move further in the right direction. On the other hand, if the second particle remains at its initial location, the first particle cannot occupy the same site due to the Pauli principle. Consequently, these two particles exhibit an effective interaction near the boundary between their dynamical regions \cite{lian2023quantumbreakdown}. This boundary interaction results in challenges for analytical solutions in this Krylov subspace.

\begin{figure}
     \centering
     \includegraphics[width=8cm]{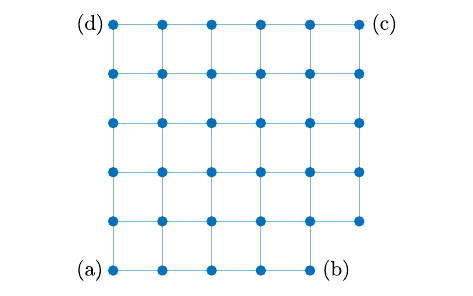}
     \caption{The connectivity of the Krylov subspace shown in Fig. \ref{fig:twoparticle}. The nodes are product states in the Krylov subspace and the edges are the nontrivial interactions induced by the Hamiltonian. The locations of configurations in Fig. \ref{fig:twoparticle} are labeled by the corresponding letters.  }
     \label{fig:two_graph}
 \end{figure}

The connectivity graph of this Krylov subspace is shown in Fig. \ref{fig:two_graph}. This graph takes the form of a square lattice, with the absence of a corner site. In the case of $J_m=J$ and $\mu_m=0$, if the distance between two fermions in the reference state [Fig. \ref{fig:twoparticle}(a)] is sufficiently large, we can ignore the boundary defect and treat the dynamics of this Krylov subspace as a free particle on a 2D square lattice. This is reasonable because the probability of finding a particle at the boundary decreases as the length of its dynamical zone increases.

  \begin{figure}
     \centering
     \includegraphics[width=8cm]{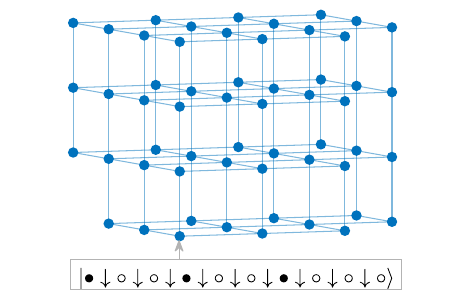}
     \caption{The connectivity of the Krylov subspace with three fermions. The reference state $\ket{\bullet\downarrow\circ\downarrow\circ\downarrow\bullet\downarrow\circ\downarrow\circ\downarrow\bullet\downarrow\circ\downarrow\circ\downarrow\circ}$ is shown below the graph.}
     \label{fig:three_graph}
 \end{figure}

The picture based on the subspace connectivity can be generalized to situations involving generically $n$ fermions. A representative connectivity graph for the subspace with $n=3$ fermions is shown in Fig. \ref{fig:three_graph}. When the fermion density in the system is sufficiently small, the average distance between two adjacent particles in the reference state like Fig. \ref{fig:twoparticle}(a) becomes significantly large. As a result, the particles have a neglected probability of simultaneously appearing at the boundary between their dynamical regions. In this context, such a boundary interaction may be considered weak and, therefore, can be neglected. Consequently, the effective dynamics can be viewed as a free particle moving on a $n$-dimensional hypercubic lattice.

\subsection{The intermediate cases}
Consider increasing the fermion density in the reference state. While ensuring that the density remains below half-filling, an increase in fermion density leads to a decrease in the average distance between two adjacent particles. Roughly speaking, the decrease in particle separation leads to strong contact interactions between neighboring particles. Consequently, the Krylov subspace becomes strongly interacting, and the dynamics dramatically deviate from those of a nearly free particle moving on a hypercubic lattice. 

\begin{figure}[t]
    \centering
    \includegraphics[width=8cm]{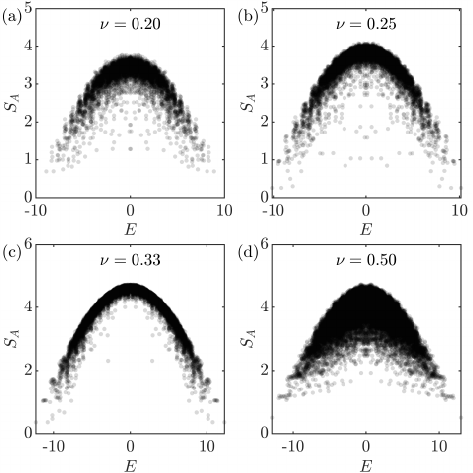}
    \caption{The subsystem entanglement entropy of eigenstates with many fermions. With the Hamiltonian in Eq.\eqref{eq:link_Ham}, we set $J_m=J=1$ and $\mu_m=0$. In each plot, $\nu=N_1/M$ denotes the filling factor of fermions, where $N_1$ is the fermion number and $M$ is the number of fermion sites. (a) $N_1=5$ and $M=25$. The gauge configuration for $\hat G_m$ in Eq. \eqref{eq:gauge_link} is set to $1000010000100001000010000$. The entanglement entropy is evaluated for subsystem A which includes the leftmost 13 fermion sites and 12 link spins. (b) $N_1=6$ and $M=24$. The gauge configuration for $\hat G_m$ is set to $100010001000100010001000$. The entanglement entropy is evaluated for subsystem A containing the leftmost 12 fermion sites and 12 link spins. (c) $N_1=8$ and $M=24$. The gauge configuration for $\hat G_m$ is set to $100100100100100100100100$. The entanglement entropy is evaluated for subsystem A which includes the leftmost 12 fermion sites and 12 link spins. (d) $N_1=11$ and $M=22$. The gauge configuration for $\hat G_m$ is set to $1010101010101010101010$,  similar to Fig. \ref{fig:PXP}. The entanglement entropy is evaluated for subsystem A that includes the leftmost 11 fermion sites and 11 link spins. }
    \label{fig:density}
\end{figure}

It is very difficult to develop an analytical description for Krylov subspaces with high densities. Therefore, we perform numerical investigations on these Krylov subspaces here. For simplicity, we focus on the situations where $J_m=J=1$ and $\mu_m=0$. Fig. \ref{fig:density} shows the subsystem entanglement entropy of the eigenstates. With an eigenstate $\ket{\psi}$, we define $\rho_A=\operatorname{Tr}_{\bar A}[\ket{\psi}\bra{\psi}]$ as the reduced density matrix of the subsystem A, where $\bar A$ is the complementary set of $A$. Then the subsystem entanglement entropy is obtained by $S_A=-\operatorname{Tr}[\rho_A\ln\rho_A]$. The numerical findings indicate that Krylov subspaces with many fermions exhibit a nonchaotic feature, as evidenced by many low-entangled eigenstates located in the central region of the spectrum. This property is linked to the boundary interactions between adjacent fermions in this kinetically constrained model.  

\subsection{PXP model and quantum many-body scar}
\begin{figure}
    \centering
    \includegraphics[width=8cm]{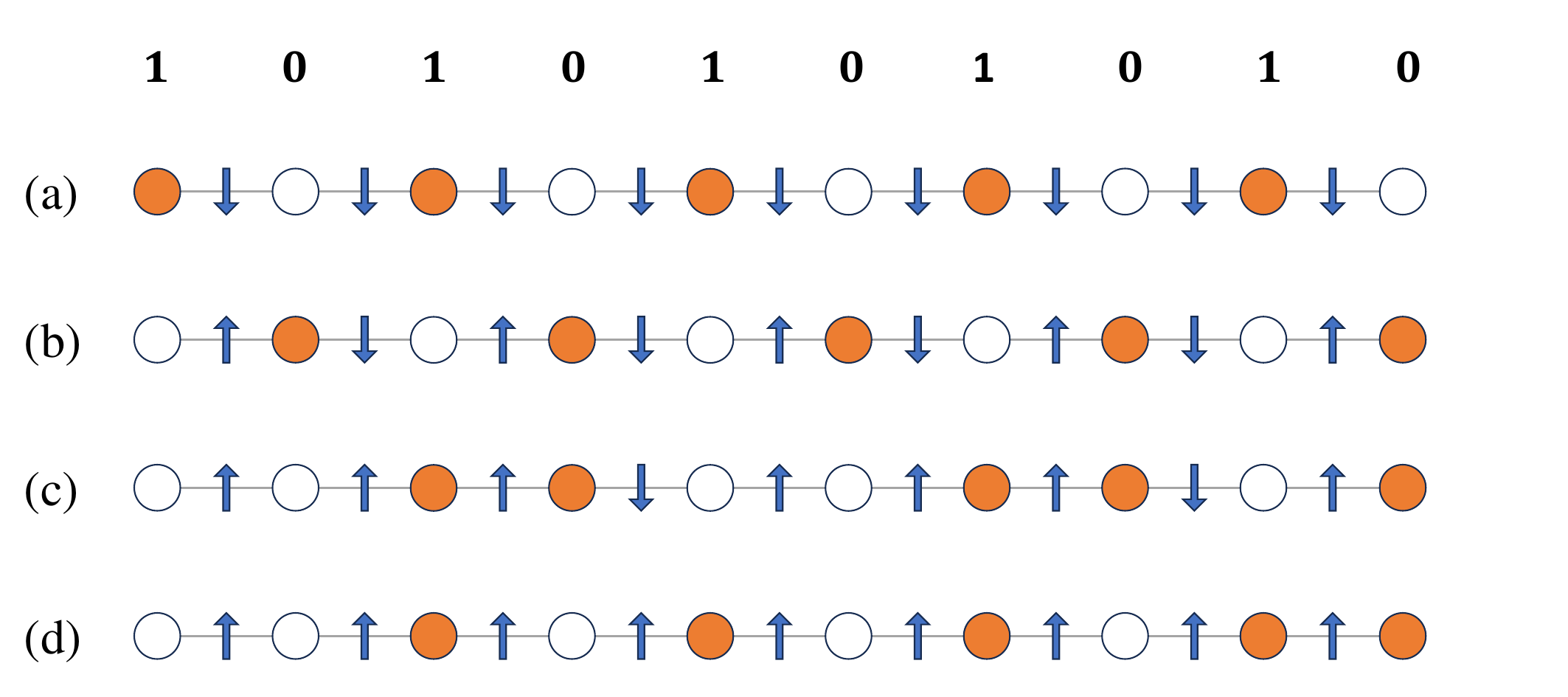}
    \caption{Typical configurations of the minimal quantum breakdown model in \cref{eq:link_Ham} in the half-filling Krylov subspace given by \cref{eq:PXP-gauge}, which maps to the PXP model. The integers above show the eigenvalues of local conserved $\hat G_m$.}
    \label{fig:PXP}
\end{figure}

\begin{figure}
    \centering
    \includegraphics[width=8cm]{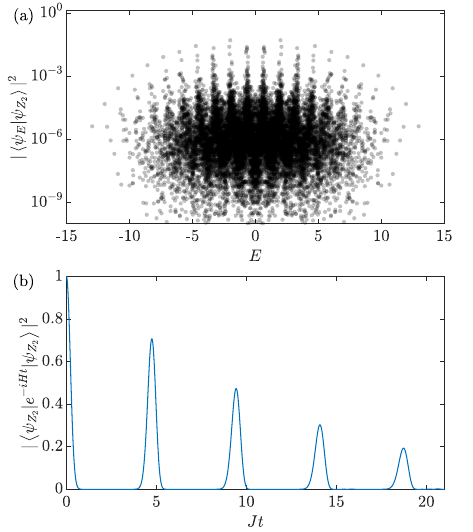}
    \caption{Quantum many-body scars in the half-filling Krylov subspace. (a) Overlap of the $Z_2$ configuration $\ket{\psi_{Z_2}}$ like Fig.\ref{fig:PXP}(a) with the energy eigenstates in this subspace. (b) The fidelity dynamics starting from $\ket{\psi_{Z_2}}$. The parameters are $J_m=J=1,\mu_m=0$, and  the number of fermionic sites is $L=22$. We take the open boundary condition in the numerical simulation. }
    \label{fig:scar_dynamics}
\end{figure}

An interesting Krylov subspace with a high density of fermions is illustrated in Fig. \ref{fig:PXP}. This corresponds to the half-filling case, where the reference state has fermions at odd sites. Within the Krylov subspace generated by this reference state, we can introduce a new gauge generator:
\begin{equation}\label{eq:PXP_generator}
    \widetilde G_m=\hat G_m-\frac{1-(-1)^m}{2}.
\end{equation}
Then all the states in this subspace satisfy
\begin{equation}\label{eq:PXP-gauge}
    \widetilde G_m\ket{\psi}=0.
\end{equation}
This precisely corresponds to the physical gauge sector discussed in lattice gauge theory for staggered fermions \cite{wiese2013ultracold,Banerjee2012atomic,Hauke2013quantum}.

Remarkably, despite the strong interactions within this subspace, the dynamics starting from the reference state reveal persistent revivals, a distinctive feature of quantum many-body scars \cite{su2023observation, Desaules2023prominent, Desaules2023weak, Halimeh2023robustquantummany,Surace2020lattice}. Notably, the Hamiltonian in this subspace of the quantum link model can be precisely mapped to the PXP model \cite{Surace2020lattice, Pan2022Composite}, a system simulated with Rydberg atoms \cite{bernien2017probing,turner2018weak,turner2018quantum}. As shown in \cref{fig:PXP}, if we map our spins on the links in \cref{eq:link_Ham} into Pauli matrices $\hat\tau^{x,y,z}_m$ through\cite{Surace2020lattice}

\begin{equation}
\begin{aligned}
    &\hat\tau^z_m\leftrightarrow(-1)^{m}\hat\sigma^z_{m,m+1},\\
   & \hat\tau^x_m\leftrightarrow\hat f_{m+1,1}^\dagger\hat \sigma^+_{m,m+1}\hat f_{m,1}+\hat f_{m,1}^\dagger\hat \sigma^-_{m,m+1}\hat f_{m+1,1},\\
   &\hat\tau^y_m\leftrightarrow -i(-1)^{m}(\hat f_{m+1,1}^\dagger\hat \sigma^+_{m,m+1}\hat f_{m,1}-\hat f_{m,1}^\dagger\hat \sigma^-_{m,m+1}\hat f_{m+1,1}).
    \end{aligned}
\end{equation}
With $J_m=J$ and $\mu_m=0$, the model \cref{eq:link_Ham} in the gauge sector of \cref{eq:PXP-gauge} maps to the PXP Hamiltonian for Rydberg atoms: 
\begin{equation}
\hat{H}_{\text{PXP}}=\frac{J}{4}\sum_{m}(1-\hat{\tau}^z_{m-1})\hat{\tau}^x_{m}(1-\hat{\tau}^z_{m+1})\ .
\end{equation}
This mapping can be understood intuitively. With the identification $\hat \sigma_{m,m+1}^z\leftrightarrow (-1)^m\hat\tau_m^z$, the bulk gauge symmetry generators \cref{eq:PXP_generator} become $\widetilde G_m=\hat f_{m,1}^\dagger\hat f_{m,1}+\frac{1}{2}[(-1)^m(\hat\tau_m^z+\hat\tau_{m-1}^z+1)-1]$, where $1<m<L$.  With $\ket{\psi}$ in the constrained Hilbert space being a basis vector, the gauge constraint \cref{eq:PXP-gauge}  leads to $\braket{\psi|\hat\tau_m^z+\hat\tau_{m-1}^z|\psi}=(-1)^m[1-2\braket{\psi|\hat f_{m,1}^\dagger\hat f_{m,1}|\psi}]-1<2$, which corresponds to the Rydberg blockade that two neighboring atoms cannot be excited simultaneously to Rydberg states. Consequently, we anticipate observing many-body scar dynamics when evolving a $Z_2$ configuration in Fig. \ref{fig:PXP}(a). 

In Fig. \ref{fig:scar_dynamics}(a), we show the overlap between the ${Z}_2$ configuration and the energy eigenstates in the subspace under the OBC. Our numerical findings distinctly reveal the presence of scar tower structures. Additionally, the fidelity $F(t) = |\bra{\psi_{Z_2}}e^{-iHt}\ket{\psi_{Z_2}}|^2$ during the time evolution exhibits clear and notable revivals, as shown in Fig. \ref{fig:scar_dynamics}(b). These outcomes collectively support the existence of quantum many-body scar states within the quantum link model, i.e., the minimal quantum breakdown model.

\section{Discussions}\label{sec:discussion}

In this paper, we study the minimal quantum breakdown model with $N=2q+1$, and investigate its understanding from the perspective of lattice gauge theory. An extensive number (proportional to the system size) of locally conserved quantities in the model leads to the emergence of Hilbert space fragmentation, which is closely tied to numerous dynamically blocking sites. The mapping between the minimal quantum breakdown model and the $U(1)$ lattice gauge theory offers a powerful framework for employing gauge configurations to describe the various dynamics within the connected Krylov subspaces of the original quantum breakdown model. These intrinsic connections among the minimal quantum breakdown model, quantum link model, and lattice gauge theory may motivate proposals for experimentally implementing the generic quantum breakdown model and more quantum models with generic modulated symmetries.

\section*{Acknowledgments}
We thank Bo-Ting Chen and Hui Zhai for helpful discussions. 

\section*{Funding}
Y.-M. H. acknowledges the support from NSFC under
Grant No. 12125405 and the Tsinghua Visiting Doctoral Students Foundation. B. L. is supported by the National Science Foundation through Princeton University’s Materials Research Science and Engineering Center DMR-2011750, and the National Science Foundation under award DMR-2141966. Additional support is provided by the Gordon and Betty Moore Foundation through Grant GBMF8685 towards the Princeton theory program.
\section*{Authors’ contributions}
All authors contributed to the theoretical study and prepared the manuscript.
\section*{Availability of data and materials  }
The data are available upon request to the authors.

\section*{Declarations}
\textbf{Competing interests.}
The authors declare that they have no competing interests.

\bibliography{ref}
\end{document}